\documentclass{xpkas}

\def\year{2015} 
\def\month{January} 
\def\volume{00} 
\def\issue{0} 
\def\beginpage{1} 
\def\endpage{99} 
\def\received{October 31, 2014} 
\def\accepted{November 30, 2014} 
\date{Received \received ; accepted \accepted}

\makeatletter

\newcommand{\Rmnum}[1]{\expandafter\@slowromancap\romannumeral #1@}
\makeatother

\def\hi{H{\scriptsize \Rmnum{1}}}

\def\cm2{cm$^{-2}$}

\def\kms{kms$^{-1}$}
\def\s{s$^{-1}$}
\def\nh3{NH$_3$}
\def\n2h{N$_2$H$^+$}
\def\co{$^{12}$CO}
\def\13co{$^{13}$CO}
\def\c18o{C$^{18}$O}
\def\hc3n{HC$_3$N}
\def\h2{H$_2$}
\def\nh{n(H$_2$)}
\def\cp{C$^+$}
\def\lp{\>\> .}
\def\lc{\>\> ,}

\def\mic{$\mu$m}

\def\aj{AJ}

\def\aap{A\&A}
\def\apjs{ApJS}
\def\nat{Nature}
\def\apj{ApJ}

\title{
Quantifying Dark Gas
}

\author[1,2]{Di Li}
\author[1]{Duo Xu}
\author[3]{Carl Heiles}
\author[1, 2]{Zhichen Pan}
\author[1]{Ningyu Tang}

\affil[1]{National Astronomical Observatories, Chinese Academy of Sciences, A20 Datun Road, Chaoyang District, Beijing 100012, China; \email{dili@nao.cas.cn , xuduo117@bao.ac.cn}}
\affil[2]{Key Laboratory for Radio Astronomy, Chinese Academy of Sciences, Nanjing 210008, China}
\affil[3]{Astronomy Department, University of California, Berkeley, CA 94720-3411}

\def\corrauthor{Di Li 

}

\def\runningauthor{
Di Li
}

\def\runningtitle{
Quantifying Dark Gas
}

\def\keywords{
ISM: dark gas; molecular gas; atomic gas; radio line
}

\def\abstracttext{
A growing body of evidence has been supporting the existence of so-called ``dark molecular gas" (DMG), which is invisible in the most common tracer of molecular gas, i.e., CO rotational emission. DMG is believed to be the main gas component of the 
intermediate extinction region between Av$\sim$0.05--2, roughly corresponding to the self-shielding threshold of \h2\ and \13co. 
To quantify DMG relative to H{\small \Rmnum{1}} and CO, we are pursuing three observational techniques, namely, \hi\ self-absorption, OH absorption, and TeraHz \cp\ emission. In this paper, we focus on preliminary results from a CO and OH absorption survey of DMG candidates.
Our analysis show that the OH excitation temperature is close to that of the Galactic continuum background and that OH is a good DMG tracer  co-existing with  molecular hydrogen in regions without CO. Through systematic ``absorption mapping" by Square Kilometer Array (SKA) and ALMA, we will have unprecedented, comprehensive knowledge of the ISM components including DMG in terms of their temperature and density, which will impact our understanding of galaxy evolution and star formation profoundly.
}

\begin{document}
\pkashead 

\section{Introduction}
 The two relatively denser phases of the interstellar medium (ISM) are the atomic Cold Neutral Medium (CNM) traced by the \hi\ $\lambda$21cm hyperfine structure line and the `standard' molecular clouds (H$_2$) as traced by CO. CO is the most important tracer of molecular hydrogen, which remains largely invisible due to lack of emission in the temperature range of molecular ISM. Empirically, CO intensities have been used as an indicator of the total molecular mass in the Milkyway and in galaxies through the so-called ``X-factor" with numerous well-known caveats.  Gases in these two phases dominate the masses of star forming clouds on a galactic scale. The measured ISM gas mass from \hi\ and CO is the foundation of many key quantities in understanding galaxy evolution and star formation, such as the star formation efficiency.
 
 A growing body of evidence, however,  indicating the existence of gas traced by neither \hi\ nor CO.  Comparative studies (e.g.\ de Vries et al.\ 1987) of Infrared Astronomy Satellite (IRAS) dust images and \hi\ and CO gas maps revealed apparent `excess' of dust emission. The Planck collaboration (2011) clearly show excess dust opacity (Fig.1) in the intermediate extinction range Av$\sim$ 0.05--2, roughly corresponding to the self-shielding threshold of \h2\ and \13co. The missing gas, or rather, the undetected gas component is widely referred to as dark gas, popularized as a common term by Grenier et al. (2005). They found more diffuse gamma-ray emission observed by Energetic Gamma Ray Experiment Telescope (EGRET) than what can be explained by cosmic-ray H-nuclei interaction (H+X-factor*CO). Observation of the TeraHz fine structure \cp\ line also help reveal dark gas in that the \cp\ line strength in diffuse gas is stronger than what can be produced by collisional excitation with only \hi\ gas (Langer et al.\ 2010).
 
 A minority of the ISM communities argued that dark gas can be explained by underestimation of HI opacities (Fukui et al.\ 2014). Due to the limited scope of this paper, we will only discuss the dark molecular gas (DMG) hereafter, or more specifically CO-dark molecular gas.
 \begin{figure}
\includegraphics[width=0.95\linewidth]{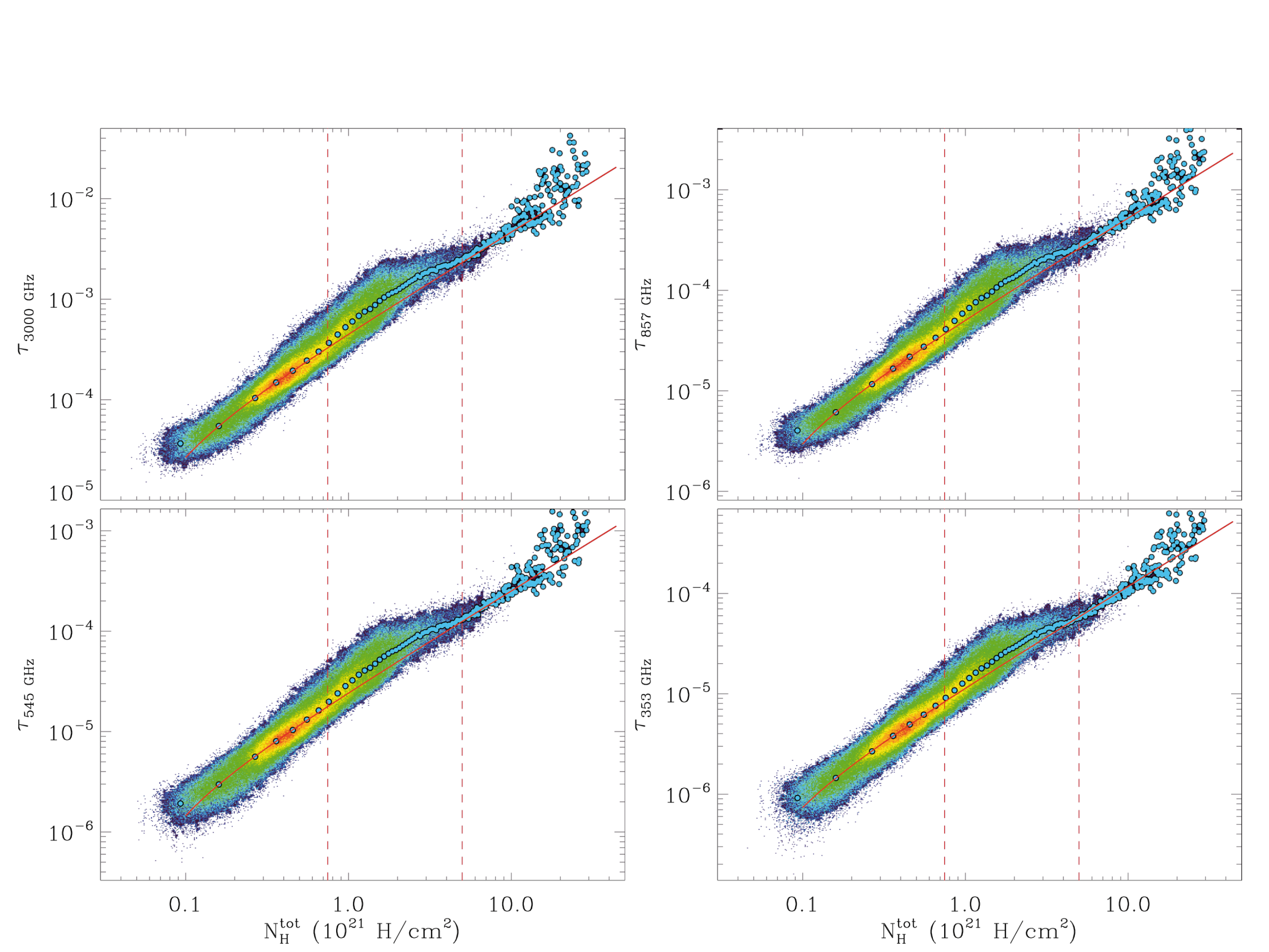}
\caption{This figure is adopted from the Figure 6 in Planck (2011). Correlation plots between the gas column density as traced by [HI+X$_{CO}$*CO] and dust optical depth at IRAS 100 \mic\ (upper left), HFI 857 GHz (upper right), 545 GHz (lower left) and 353 GHz (lower right). The color scale represents the density of sky pixels on a log scale. The blue dots show a binned average representation of the correlation. The red line shows the best linear correlation derived at low values. The vertical lines show the positions corresponding to Av = 0.37 mag and Av = 2.5 mag.  A single CO X-factor X$_{CO}$ = 2.3 $\times10^{20}$ \h2\ cm$^{-2}$/(K km \s) was used. }
\label{fig1}
\end{figure}

 It is natural to infer from chemical and PDR models that molecular hydrogen would exist in regions where CO is not detectable.
 CO can be of  low abundance due to photo-dissociation (Fig.2) in unshielded regions and/or can be heavily sub-thermal due to lack of collision in diffuse gas. We strive to provide direct measurements and/or constraints of the physical conditions of DMG. Section 2 will focus on OH absorption. Section 3 introduces a CO survey toward background continuum sources. Section 4 presents the combined analysis of CO and OH followed by a brief outlook of upcoming surveys in the final section.

 \section{Where is the Hydroxyl?}
 
 OH, or Hydroxyl, is the first interstellar molecule detected in radio bands (Weinreb et al.\ 1963). It can form quickly through a series of charge exchange reactions initiated by cosmic ray once \h2\ is present (van Dishoeck \& Black 1988). OH can also form on grains. One of the main chemical paths associated with CO after OH formation is
 \begin{equation}
\mathrm{OH+C^+\to CO^++ H}\label{eq1} \lc
\end{equation}
\begin{equation}
\mathrm{CO^++H_2 \to HCO^++ H}\label{eq2} \lc
\end{equation}
\begin{equation}
\mathrm{HCO^++e^-\to CO+ H}\label{eq3}  \lp
\end{equation}

We should expect wide-spread and abundant OH along with HCO$^+$ and \cp. HCO$^+$ is accessible in millimeter bands. The main transition from \cp\ is its fine structure transition in 2 THz band,  which is impossible to map from the ground. It is somewhat puzzling why large scale OH surveys of ISM has not been available in the half a century since its discovery. In fact, thousands of hours of Arecibo time have been spent on searches for OH in galaxies with mostly negative results (e.g.\ Schmelz \& Baan 1988).

In contrast, Dickey et al.\ (1981) found OH in absorption against high Galactic latitude continuum sources. Important and extensive confirmatory absorption measurements by Liszt \& Lucas(1996) and Lucas \& Liszt(1996) found that OH and HCO$^{+}$ are commonly observed against such sources. Lucas \& Liszt(1996) found that $\thicksim$30\% of continuum sources having \hi\ in absorption exhibit HCO$^{+}$ in absorption. In light these results, the dearth of OH emission should be attributed to the excitation condition of OH rather than its abundance.
 
 The observed antenna temperature $T_A$ is 
 \begin{equation}
T_A=(T_{ex}-T_{C})(1-e^{-\tau})
\label{eq4} 
\end{equation}
where $T_C \sim 3.5$ K is the continuum background temperature at L-band, comprising of the CMB and the galactic synchrotron emission. 
When $T_{ex}$ approaches $T_C$, the apparent signal from certain line emission vanishes. Such gas, however, is suitable for absorption study when the telescope is trained toward background sources with $T_C \gg T_{ex}$ as is the case when observing quasars and/or HII regions.

Heiles \& Troland (2003) published the well-cited Millennium survey of 21-cm line absorption toward 79 continuum sources. The ON-OFF technique and Gaussian decomposition analysis allow them to provide credible measurements of the excitation temperature and density of \hi\ components spreading through the Milkyway. Unpublished OH absorption data were taken simultaneously with the Millennium survey. Our preliminary analysis of these OH absorption data confirms the suspicion that $T_{ex}$ of OH aggregates around the background temperature, and thus renders OH undetectable in emission.

\begin{figure}
\includegraphics[width=0.96\linewidth]{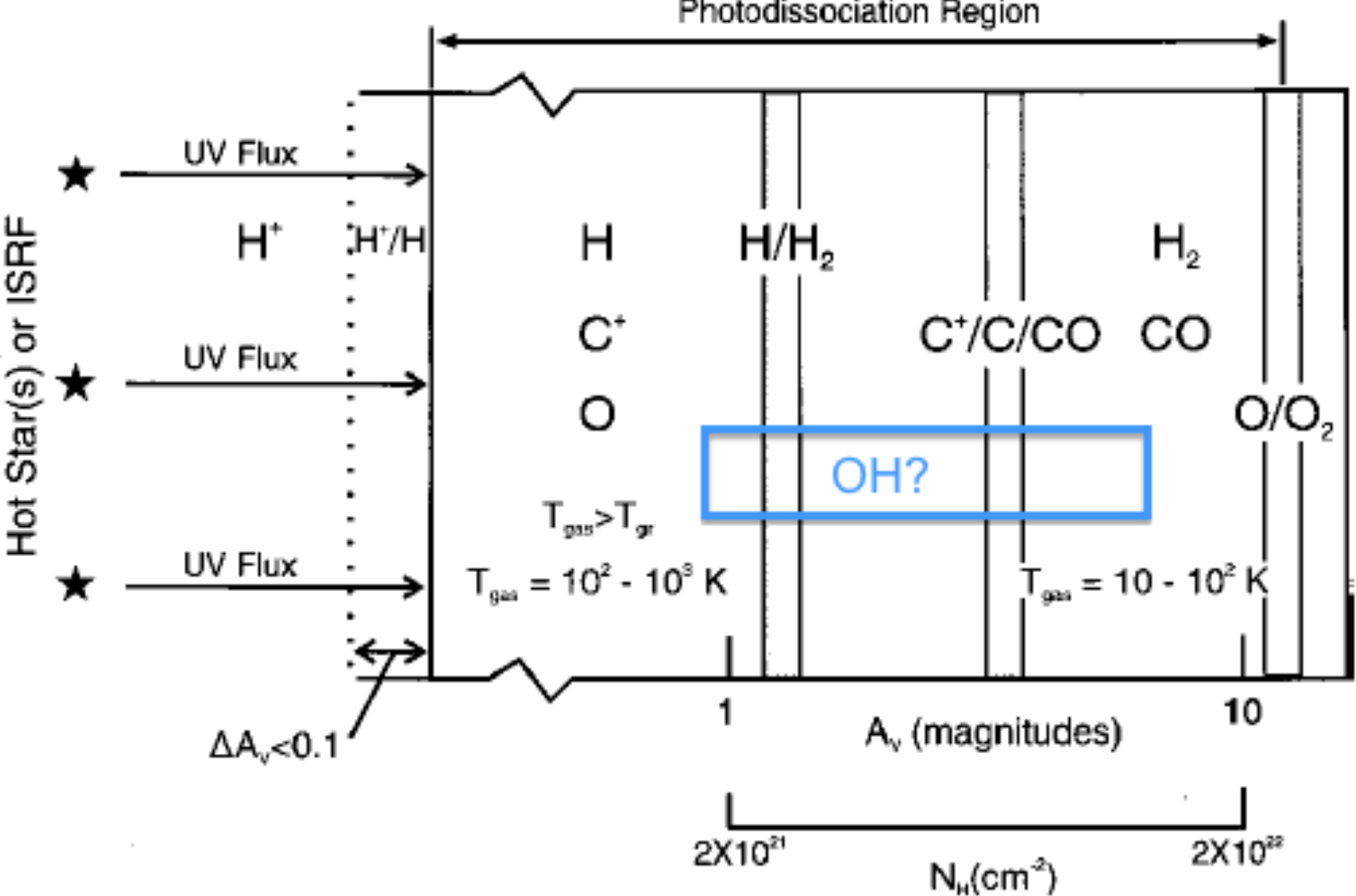}
\caption{A schematic view of photo-dissociation region (Tielens  2005). It shows the locations of different transition layers. We add the blue pane to indicate the possible location of OH.}
\label{pdrmodel}
\end{figure}

 \section{A Multi-transition CO Survey of Millennium Sources}
	We conducted a follow-up CO survey of all Millennium sight-lines with OH absorption. 
	Toward 79 published Millennium survey sources, 43 sight-lines exhibit OH absorption. 
	These 43 sources with OH absorption have been observed in \co\  J=1-0 and 2-1, \13co\  J=1-0 and \c18o\  J=1-0. 
	The 8 sources with strong \co\  J=1-0 and 2-1 lines are also observed in the \co\  J=3-2 transition.
	
	The J=1-0 transition of $^{12}$CO,  $^{13}$CO and \c18o\  were observed in March and May of 2013 and May of 2014 with
    the Purple Mountain Observatory Delingha (PMODLH) 13.7 m telescope, Chinese Academy of Sciences. 
    The spectra was obtained with position switch mode, the reference position is selected from IRAS Sky Survey Atlas\footnote{$http://irsa.ipac.caltech.edu/data/ISSA/$}. 
	With 1000 MHZ bandwidth spectroscopy, the frequency resolution is 61 kHz, results in an approximate 0.18 \kms channel width. 
	
	The \co\  J=2-1 and J=3-2 data were taken with the Caltech Submillimeter Observatory (CSO) 10.4m on top of Mauna Kea in July, October and December of 2013. 
	The velocity resolution for $^{12}$CO(J=2-1) spectra is 0.16 \kms. 
	The velocity resolution for $^{12}$CO(J=3-2) spectra is 0.11 \kms or 0.56 \kms due to a problem of spectroscopy. 

\begin{figure}
\includegraphics[width=1.0\linewidth]{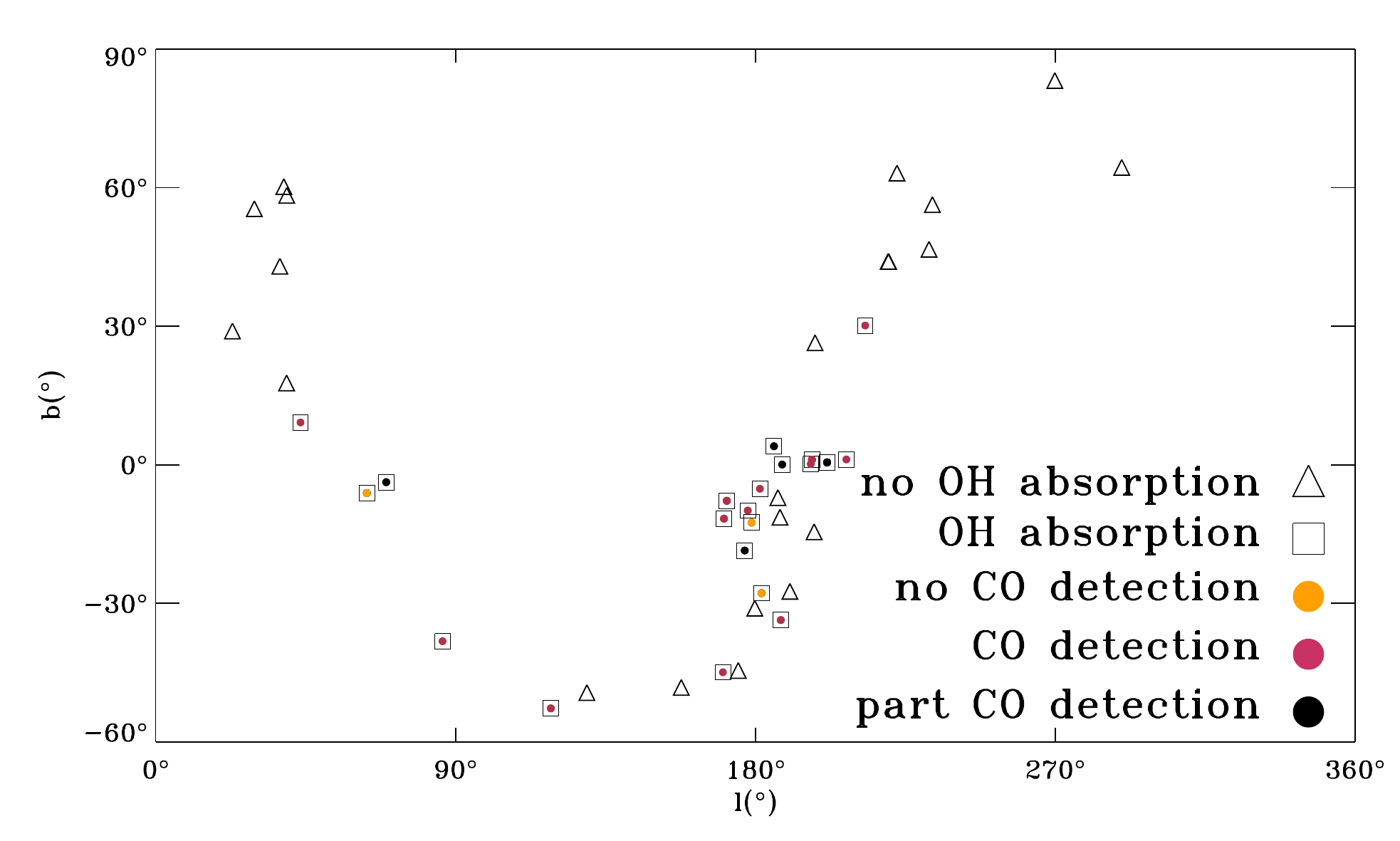}
\caption{The location of sources in galactic coordinate. Triangle represents the source without any absorption component in OH. Square represents the source with absorption component in OH. Yellow dot represents the source without any detection of CO transitions. Red dot represents the source with CO transition components, and these CO components correspond to all the absorption components in OH. Black dot represents the source with CO transition components, and these CO components correspond to some of the absorption components in OH, but there are no detectable CO transition corresponding to the rest of absorption components in OH. We call these kind of sources ``part CO detection".  }
\label{fig2} 
\end{figure}

	The distribution of these spectra in the galactic coordinate is presented in Fig.~\ref{fig2}.  
	
	The typical RMS of \co\  J=1-0 observation is about 0.06K, which corresponds to a CO detection limit of 2.6$\times$10$^{13}$ cm$^{-2}$.

    The astronomical software package Gildas/CLASS\footnote{$http://www.iram.fr/IRAMFR/GILDAS/$} was used for baseline removing, spectra combining and gaussian fitting. 

\section{Comparison of \hi, OH, and CO}
\begin{figure}[htp]
\includegraphics[width=1.0\linewidth]{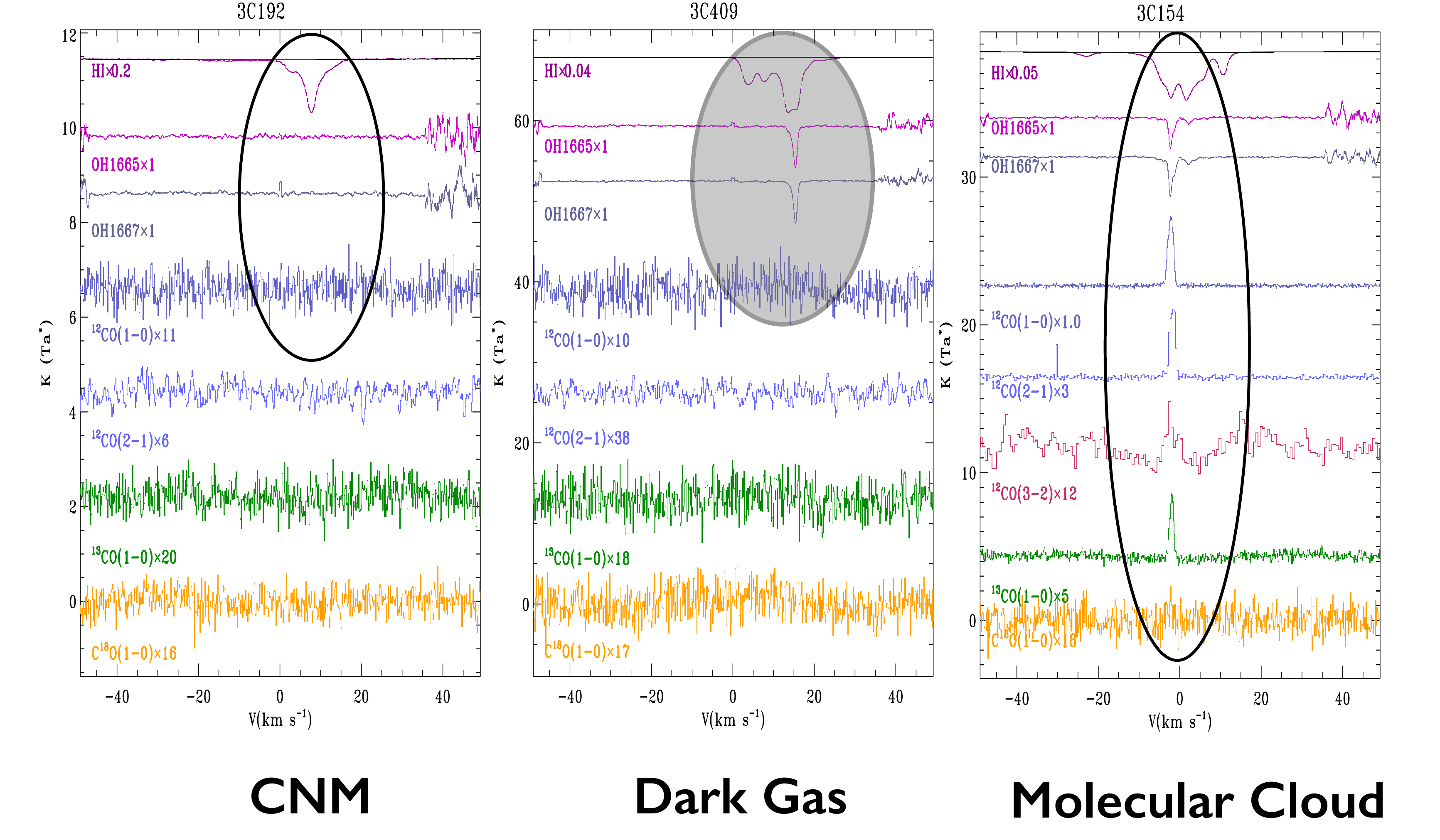}
\caption{Representative spectra. 3C192 sightline has only \hi\ seen in absorption. One component of 3C154 sightline has OH and \hi\ in absorption and CO and its isotopologues in emission. 3C409 sightline has one gas component with both OH and \hi\ , but no detectable CO transitions.}
\label{fig4}
\end{figure}

We compare the Gaussian components seen in \hi\ absorption, OH absorption, and CO emission.
A total of 115 Gaussian components were detected as specified in Heiles \& Troland (2003). 
52 such gas components have OH absorption. The majority of these 52  have CO emission, except for 13 components, which are
DMG candidates. There are no component with  only CO emission and no OH absorption.
Three representative sets of spectra are shown in Fig.~\ref{fig4}.  Toward 3C192, only \hi\ is present, typical of CNM. 
Toward 3C154, there is a  component with \hi, OH, and several CO and CO isotopologue transitions, which should be 
representative of `normal' molecular clouds. Toward 3C409, there exists a component with \hi\ and OH absorption, but no CO emission in any
transition observed. We posit that this is typical of DMG.  The percentage of these three categories are 55\% CNM,  34\% molecular clouds, and 11.3\% DMG.

The column densities of  OH components were calculated as
\begin{equation}
N_{OH}=\frac{8\pi kT_{ex}{{\nu}^2_{1667}}}{A_{1667}c^{3}h}\frac{16}{5}\int \tau_{1667} \,dv \lc
\label{eq5}
\end{equation}
where $A_{1667}=7.778\times 10^{-11} s^{-1}$ is the A-coefficient  and $T_{ex}$ is its excitation temperature calculated based on a
recipe similar to that for \hi\ absorption components in Heiles \& Troland (2003). 

The CO column densities were calculated in two categories. If only the J=1-0 transition of \co\ is detected, the optical depth is assumed to be
small and the excitation temperature is assumed to be the same as that of OH. If both \co\ and \13co\ are detected, we derive the optical depth and the excitation temperature based on multiple transitions and Local Thermodynamic Equilibrium (LTE) assumptions. The recipe for deriving CO column densities can be found in
Li (2002).

\begin{figure}
\includegraphics[width=1.0\linewidth]{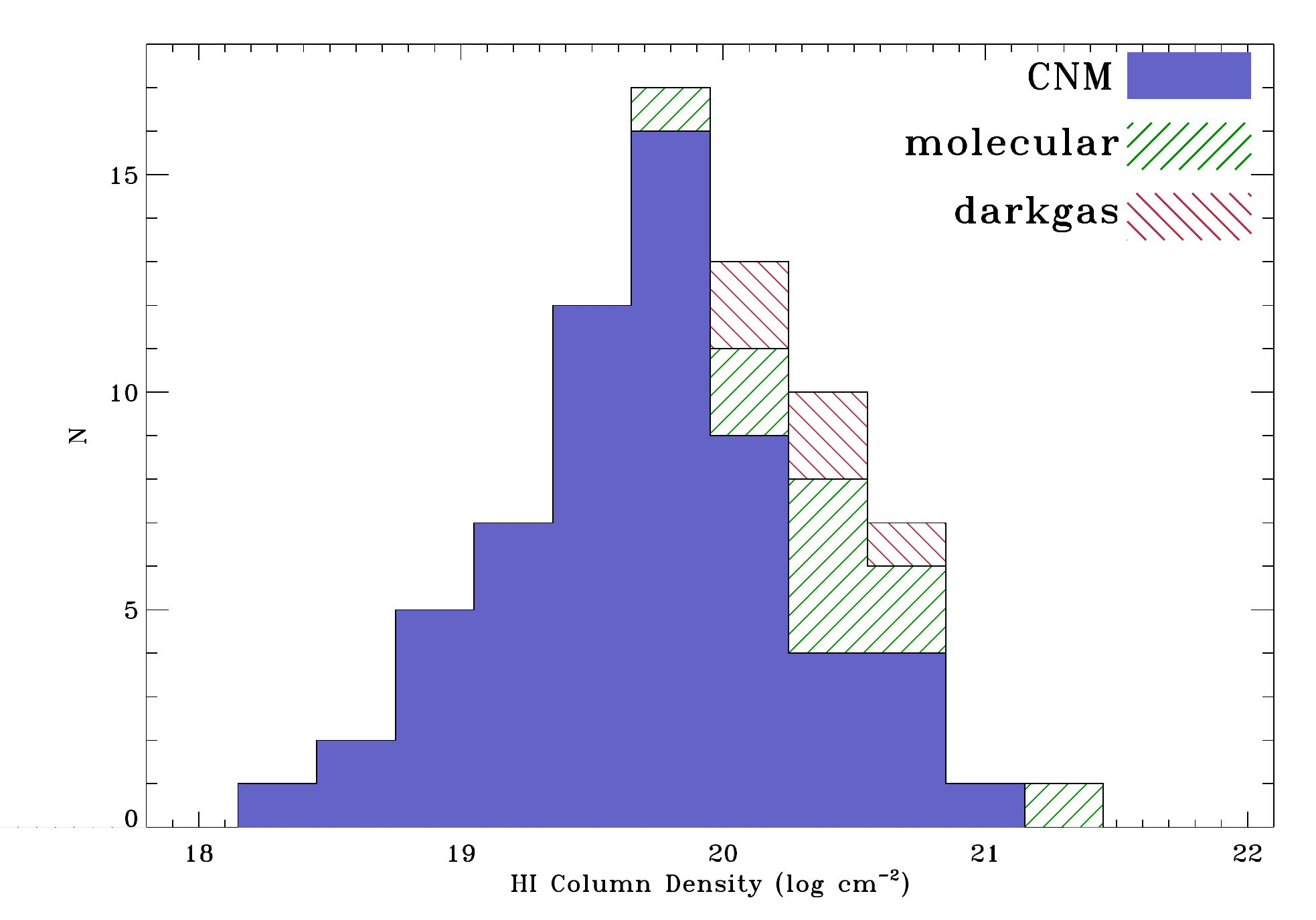}
\caption{The histogram of \hi\ colunm density for the cold neutron medium(the blue filled rectangle), molecular gas(the green line filled rectangle) and dark gas(the red line filled rectangle).  }
\label{fig5}
\end{figure}

The statistics of gas column densities (Fig.~\ref{fig5}) is consistent with the schematic picture presented in Fig.~\ref{fig2} and section 1.
There is an apparent gas column density threshold for OH detection at around Av$\sim$0.05, above which OH and CO have similar distribution.
OH turns out be a good tracer of diffuse gas with `intermediate' extinction, namely, between the self-shielding threshold for \h2\ and \13co.

\section{Discussion}
The expected location, abundance, and optical depth of OH should make it 
an excellent tracer of DMG. Due to insufficient collision in diffuse gas, however, OH is hard to
detect in emission. This is likely the main reason why a galactic scale or even any large-scale OH map has not 
been accomplished. The realize its potential in quantifying dark gas throughout the ISM, the upcoming radio telescopes 
will be needed to conduct comprehensive absorption surveys. The Five-hundred-meter Aperture Spherical radio Telescope (FAST)
is expected to start operation in late 2016. The unprecedented sensitivity of FAST and its early science instruments (Li et al.\ 2013) 
should make feasible a \hi+OH absorption survey, in the mode of the Millennium survey, but with 10 times more sources. 
The SKA1 will have the survey speed and sensitivity to measure gas absorption with a source density between a few to a few tens per square degree (McClure-Griffiths et al.\ 2014), which means that an all sky ``absorption-image" is feasible and we will have ISM temperature and density 
everywhere! Based on similar excitation  and sensitivity considerations,  ALMA is a powerful instrument to obtain systematic and sensitive
absorption measurements of millimeter lines in diffuse gas. CO and HCO$^+$ in diffuse gas, in particular, will be much better constrained in terms of excitation temperature and column densities through ALMA absorption observation than emission measurements. Combining both radio and millimeter absorption surveys in the coming decade, we will quantify DMG and provide definitive answers to questions like the
global star formation efficiency. 

This work is supported in part by National Basic Research Program of China (973 program) No. 2012CB821800, NSFC No. 11373038, and  Chinese Academy of Sciences Grant No. XDB09000000.
We thank Lei Qian for his great help in running $^{12}$CO J=3-2 observation in CSO site. 
We also thank Lei Zhu for his help in CSO observation. 
The 3$\times$3 multibeam sideband separation superconductor-insulator-superconductor (SIS) receiver with sideband separation (Zuo et al. 2011, Shan et al. 2012) was used for observation. The \co\ ,\13co\  and \c18o\  spectra was obtained with position switch mode, the reference position is selected from IRAS Sky Survey Atlas\footnote{$http://irsa.ipac.caltech.edu/data/ISSA/$}. 
The authors appreciate all the staff members of the PMODLH for their help during the observation. 
This material is partly based upon work at the Caltech Submillimeter Observatory, which is operated by the California Institute of Technology.

\end{document}